\documentclass[preprint2]{aastex631}

\usepackage{mathrsfs }
\usepackage{natbib}
\usepackage{amssymb}
\usepackage{apjfonts}
\usepackage{ulem} 
\usepackage{graphicx}
\begin{document}
\clearpage

\title{The MOSDEF Survey: Implications of the Lack of Evolution in the Dust Attenuation-Mass Relation to $\lowercase{z}\sim 2$\footnote{Based on data obtained at the W.M. Keck Observatory, which is operated as a scientific partnership among the California Institute of Technology, the University of California,  and the National Aeronautics and Space Administration, and was made possible by the generous financial support of the W.M. Keck Foundation.}}

\author{Alice E. Shapley}
\affiliation{Department of Physics and Astronomy, University of California, Los Angeles, 430 Portola Plaza, Los Angeles, CA 90095, USA}

\author{Ryan L. Sanders}
\affiliation{Department of Physics, University of California, Davis, 1 Shields Avenue, Davis, CA 95616, USA}
\affiliation{Hubble Fellow}

\author{Samir Salim}
\affiliation{Department of Astronomy, Indiana University, Bloomington, IN 47404, USA}

\author{Naveen A. Reddy}
\affiliation{Department of Physics and Astronomy, University of California, Riverside, 900 University Avenue, Riverside, CA 92521, USA}

\author{Mariska Kriek}
\affiliation{Astronomy Department, University of California at Berkeley, Berkeley, CA 94720, USA}
\affiliation{Leiden Observatory, Leiden University, NL-2300 RA Leiden, Netherlands}

\author{Bahram Mobasher}
\affiliation{Department of Physics and Astronomy, University of California, Riverside, 900 University Avenue, Riverside, CA 92521, USA}

\author{Alison L. Coil}
\affiliation{Center for Astrophysics and Space Sciences, Department of Physics, University of California, San Diego, 9500 Gilman Drive., La Jolla, CA 92093, USA}

\author{Brian Siana}
\affiliation{Department of Physics and Astronomy, University of California, Riverside, 900 University Avenue, Riverside, CA 92521, USA}

\author{Sedona H. Price}
\affiliation{Max-Planck-Institut f\"ur Extraterrestrische Physik, Postfach 1312, Garching, 85741, Germany}

\author{Irene Shivaei}
\affiliation{Steward Observatory, University of Arizona, 933 N Cherry Ave, Tucson, AZ 85721, USA}

\author{James S. Dunlop}
\affiliation{Institute for Astronomy, University of Edinburgh, Royal Observatory, Edinburgh EH9 3HJ, UK}

\author{Ross J. McLure}
\affiliation{Institute for Astronomy, University of Edinburgh, Royal Observatory, Edinburgh EH9 3HJ, UK}

\author{Fergus Cullen}
\affiliation{Institute for Astronomy, University of Edinburgh, Royal Observatory, Edinburgh EH9 3HJ, UK}

\email{aes@astro.ucla.edu}

\shortauthors{Shapley et al.}



\shorttitle{Dust Attenuation-Mass Relation to $z\sim 2$}

\begin{abstract}
We investigate the relationship between dust attenuation and stellar mass ($M_*$)
in star-forming galaxies over cosmic time. For this analysis, we compare measurements
from the MOSFIRE Deep Evolution Field (MOSDEF) survey
at $z$$\sim$$2.3$ and the Sloan Digital Sky Survey (SDSS) at $z$$\sim$$0$, augmenting
the latter optical dataset with both UV Galaxy Evolution Explorer (GALEX) and
mid-infrared Wide-field Infrared Survey Explorer (WISE) photometry from the
GALEX-SDSS-WISE Catalog.  We quantify dust attenuation using both spectroscopic
measurements of H$\alpha$ and H$\beta$ emission lines, and photometric measurements
of the rest-UV stellar continuum. The H$\alpha$/H$\beta$ ratio is used to determine
the magnitude of attenuation at the wavelength of H$\alpha$,
$A_{{\rm H}\alpha}$. Rest-UV colors and spectral-energy-distribution fitting
are used to estimate $A_{1600}$, the magnitude of attenuation at a rest wavelength
of 1600~\AA. As in previous work, we find a lack of significant evolution
in the relation between dust attenuation and $M_*$ over the redshift
range $z$$\sim$$0$ to $z$$\sim$$2.3$. Folding in the latest estimates of the
evolution of $M_{{\rm dust}}$, $({M_{{\rm dust}}}/{M_{{\rm gas}}})$, and gas surface density at fixed $M_*$, 
we find that the expected $M_{{\rm dust}}$ and dust mass surface density are both significantly
higher at $z$$\sim$$2.3$ than at $z$$\sim$$0$. These differences appear at odds with the lack of evolution in dust attenuation. To explain the striking constancy
in attenuation vs. $M_*$, it is essential to determine the relationship between metallicity and $({M_{{\rm dust}}}/{M_{{\rm gas}}})$,
the dust mass absorption coefficient, and dust geometry, and the evolution of these relations and quantities
from $z$$\sim$$0$ to $z$$\sim$$2.3$.
\end{abstract}


\section{Introduction}
\label{sec:intro}

Tracing the effects of dust attenuation on starlight is crucial for obtaining
a complete census of star formation over cosmic time \citep[e.g.,][]{madau2014}.
There are many different methods for quantifying the role played by dust
in star-forming galaxies over a wide range in redshift. These rely on multi-wavelength
data probing the ratio of far-infrared (far-IR) to UV emission -- i.e., re-radiated
vs. unobscured starlight; and various measures of UV/optical reddening and/or attenuation.
Measuring the dust content of star-forming galaxies over cosmic time in systems
spanning a range of stellar and gas masses also places constraints on models
for the formation and destruction of dust grains in the interstellar medium (ISM) \citep[e.g.,][]{popping2017}.

One particularly striking observation regarding dust in star-forming galaxies
is that the relationship between dust attenuation and stellar mass ($M_*$) does {\it not}
significantly evolve between $z\sim 0$ and $z\sim 2$ (and perhaps to even higher
redshift). Here dust attenuation has been quantified as the ratio of far-IR to UV
star-formation rates (SFRs) or luminosities, which is also called ``IRX" \citep[e.g.,][]{meurer1999,heinis2014,bouwens2016,bourne2017};
the magnitude of far-UV (i.e., 1600\AA) attenuation, or $A_{1600}$ \citep[e.g.,][]{mclure2018,pannella2015};
the fraction of star formation that is obscured, $f_{\rm obscured}$ \citep{whitaker2017},
and the nebular attenuation based on the Balmer decrement \citep[i.e., H$\alpha$/H$\beta$ ratio;][]{kashino2013,dominguez2013,price2014}.
There is less consensus regarding the form of the attenuation vs. $M_*$ relation at $z>3$,
with some evidence that it may evolve towards lower attenuation at fixed $M_*$ \citep[e.g.,][]{fudamoto2020}.
However, at least out to $z\sim 2$, multiple results suggest a constant relation between
dust attenuation and $M_*$ in mass-complete samples \citep{whitaker2017,mclure2018}.

The degree of dust attenuation in a galaxy reflects multiple key features
of its ISM. First, there is the total dust content, $M_{\rm dust}$,
and its relationship with the gas content of the galaxy $M_{\rm gas}$. This dust content
is intimately connected with the degree of metal enrichment in the galaxy, given
that dust grains form from heavy elements \citep{remyruyer2014,devis2019}.
However, dust attenuation  reflects not only the total dust content,  $M_{\rm dust}$,
but also its spatial distribution, which can be quantified in the most simplistic manner in terms
of a characteristic radius, $r_{\rm dust}$ (i.e., a dust-continuum half-light radius). 
In more detail, the non-uniformity of the spatial distribution of dust must also be 
taken into account \citep[e.g.,][]{witt2000,charlot2000,seon2016}.
Finally, there is the
question of the very properties of the dust grains, including their chemical composition,
size distribution and morphologies, which determines the relationship between 
$M_{\rm dust}$ and opacity.

Thus far, analyses of the (lack of) evolution in the attenuation vs. $M_*$
relation have not incorporated what is known about change in these other ISM
components: $M_{\rm gas}$, metallicity ($12+\log({\rm O/H}$), and $M_{\rm dust}$.
Yet, they must be considered in order to gain a full understanding 
of why a measurement of $M_*$ at $z\sim 0$ to $z\sim 2$ is so determinative
of the degree of dust attenuation. Here, we analyze $z\sim 2.3$ dust attenuation
based on rest-optical spectroscopic measurements of the H$\alpha$/H$\beta$
Balmer decrement and rest-UV continuum measures of dust reddening. 
We also fold in independent results on the evolution of galaxy metallicities and gas
and dust content, in order to gain a complete picture of the
ISM of star-forming galaxies over the past $\sim 10$~billion years.
In \S\ref{sec:obs}, we describe
our samples and observations. In \S\ref{sec:results}, we present results on the observed
relationship between dust attenuation and stellar mass at both $z\sim 2.3$ and $z\sim 0$.
In \S\ref{sec:discussion}, we discuss the surprising implications of the lack of strong evolution
in the attenuation vs. mass relation.  Throughout, we adopt cosmological parameters of
$H_0=70\mbox{ km  s}^{-1}\mbox{ Mpc}^{-1}$, $\Omega_m=0.30$, and
$\Omega_{\Lambda}=0.7$, and a \citet{chabrier2003} IMF.

\section{Sample and Observations}
\label{sec:obs}

\subsection{MOSDEF}
\label{sec:obs-mosdef}
The analysis presented here is based on data from the MOSFIRE Deep Evolution Field (MOSDEF)
survey, a large observing program using the Multi-Object
Spectrometer for Infrared Exploration \citep[MOSFIRE;][]{mclean2012}
on the 10~m Keck~I telescope. As described in \citet{kriek2015},
with the MOSDEF survey we obtained rest-optical spectra for a sample of $\sim 1500$ galaxies
within three distinct redshift intervals spanning $1.4 \leq z \leq 3.8$. These intervals
are $1.37 \leq z \leq 1.70$, $2.09 \leq z \leq 2.61$, and $2.95 \leq z \leq 3.80$, where the strongest
rest-optical emission lines can be observed within windows of atmospheric transmission. MOSDEF targets fell in the COSMOS,
GOODS-N, AEGIS, GOODS-S, and UDS fields, in regions covered by the 
CANDELS and 3D-HST surveys \citep{grogin2011,koekemoer2011,momcheva2016}.
These fields feature extensive multi-wavelength datasets spanning the electromagnetic
spectrum, which can be used to infer a wide range
of galaxy properties. The data used for fitting the spectral energy distributions (SEDs) of
MOSDEF galaxies  have been cataloged by the 3D-HST survey \citep{skelton2014}, and include optical
and near-infrared ground-based and {\it Hubble Space Telescope} photometry, as well as {\it Spitzer}/IRAC
mid-infrared measurements. MOSDEF targets are selected based on existing photometric or spectroscopic redshifts,
with a magnitude limit in the rest-optical (observed $H$ band). This limit is $H_{AB}=$24, 24, and 25, respectively,
for the lowest, middle, and highest redshift interval of the survey.

Here we focus on MOSDEF star-forming galaxies within the central target redshift range,
i.e., $2.09 \leq z \leq 2.61$. Our $z\sim$~2 sample is very similar to the one analyzed in \citet{sanders2021},
with the additional constraint of $\geq 3\sigma$ detections of both H$\alpha$ and H$\beta$ line fluxes.
Each galaxy has a robust estimate of nebular oxygen abundance ($12+\log({\rm O/H})$) based on
the subset of detected strong nebular emission lines drawn from [OII]$\lambda\lambda3726,3729$, [NeIII]$\lambda3869$, H$\beta$, and 
[OIII]$\lambda5007$ (at the very least [OII]$\lambda\lambda3726,3729$, H$\beta$, and [OIII]$\lambda5007$), as described
in \citet{sanders2021}, and a measure of nebular dust attenuation ($A_{{\rm H}\alpha}$) based on the ratio H$\alpha$/H$\beta$
and assuming the \citet{cardelli1989} extinction law.
We also estimated stellar masses by modeling the multi-wavelength photometric spectral energy distributions (SEDs) cataloged by the
3D-HST team \citep{skelton2014}, where the near-infrared photometry was corrected for the contribution
of strong nebular emission lines \citep{sanders2021}. For SED modeling, we used the program FAST \citep{kriek2009}, assuming the stellar population synthesis models of \citet{conroy2009}, a \citet{chabrier2003} IMF, a \citet{calzetti2000} dust law, and delayed-$\tau$
star-formation histories, where $SFR(t)\propto t \exp(-t/\tau)$. Here, $t$ is the time since the onset of star formation and $\tau$ is the 
characteristic star-formation timescale. We note that Balmer emission-line fluxes were corrected for the underlying 
stellar absorption implied by the best-fit
stellar population model, with typical Balmer absorption corrections of $\sim 1$\% for H$\alpha$ and $\sim 7$\% for H$\beta$. Finally, AGNs were identified and removed based on their X-ray and infrared
properties, as well as those with $\log({\rm [NII]/H}\alpha)>-0.3$ \citep{coil2015,azadi2017,leung2019}.
In total, our MOSDEF sample includes 210 galaxies 
with a median redshift of $z_{\rm med}=2.28$, a median stellar mass of $\log(M_*/M_{\odot})_{\rm med}=9.88$, and a median
dust-corrected H$\alpha$-based SFR of ${\rm SFR}_{\rm med} = 29 M_{\odot}\mbox{ yr}^{-1}$.
These median properties are very well-matched to those of the $z\sim 2.3$ sample analyzed in \citet{sanders2021}.

As an additional measure of dust attenuation, which probes the stellar continuum, we estimated the UV slope, $\beta$, directly from broadband
photometry. $\beta$  is calculated by fitting a power
law of the form $f_{\lambda} \propto \lambda^{\beta}$ to the photometric bands spanning the rest-wavelength range $1268-2580$\AA.
This fit is typically determined based on 4$-$5 bands for galaxies in all fields except COSMOS, where the fit is typically
based on 20 bands. The values of $\beta$ were translated into estimates of rest-UV continuum attenuation ($A_{1600}$)
for our sample based on a few different prescriptions. Following \citet{reddy2018} and assuming an intrinsic stellar population from the Binary Population
and Spectral Synthesis (BPASS) code \citep{eldridge2017}, including binaries, an upper mass cut-off of $300 M_{\odot}$, $Z_*=0.02$,
and $\log({\rm age/yr})=8.0$, we find the relation 

\begin{equation}
A_{1600}=2.13\times\beta+5.04
\label{eq:a1600calz}
\end{equation}

\noindent for a \citet{calzetti2000} dust attenuation law, and 

\begin{equation}
A_{1600}=1.07\times\beta+2.52
\label{eq:a1600smc}
\end{equation}

\noindent for an SMC extinction law \citep{gordon2003}. We note that the relationship above for the \citet{calzetti2000} law is based on assuming an intrinsically bluer UV slope of $\beta_{\rm int}=-2.37$ for $z\sim 2$ star-forming galaxies than that found in earlier work for $z\sim 0$ starbursts. Specifically, in \citet{meurer1999}, the relationship between $A_{1600}$ and $\beta$ ($A_{1600}=1.99\times\beta+4.43$) assumes an intrinsic UV slope of $\beta_{\rm int}=-2.23$. We also note that if BPASS models like the ones used to derive 
equations~(\ref{eq:a1600calz}) and (\ref{eq:a1600smc}) are adopted to estimate stellar masses, we obtain results extremely consistent
with those based on FAST models. The same holds if we adopt \citet{bruzual2003}, constant star formation, 
solar-metallicity models to estimate stellar masses.

Recent results from the MOSDEF survey \citep{shivaei2020} suggest that a \citet{calzetti2000}-type curve
is appropriate at metallicities of $12+\log({\rm O/H}) \geq 8.5$, while one resembling the SMC curve seems
to apply at $12+\log({\rm O/H}) < 8.5$. These results echo previous evidence from \citet{reddy2010} and \citet{reddy2012}
that an SMC curve best describes the youngest (age$<$100~Myr) systems among a large sample of UV-selected
galaxies at $z\sim 2$. Other recent work \citep[e.g.,][]{mclure2018} presents evidence
that the \citet{calzetti2000} applies over a wide range of stellar masses ($\log(M_*/M_{\odot})\geq 9.75$), and, correspondingly, metallicities.
Accordingly, our two favored methods for translating $\beta$ into $A_{1600}$ values for MOSDEF galaxies consist
of (1) assuming the \citet{calzetti2000} curve for the entire sample; (2) assuming the \citet{calzetti2000} curve
at $12+\log({\rm O/H}) \geq 8.5$ and the SMC curve at  $12+\log({\rm O/H}) < 8.5$. 

\begin{figure*}[t!]
\centering
\includegraphics[width=0.95\linewidth]{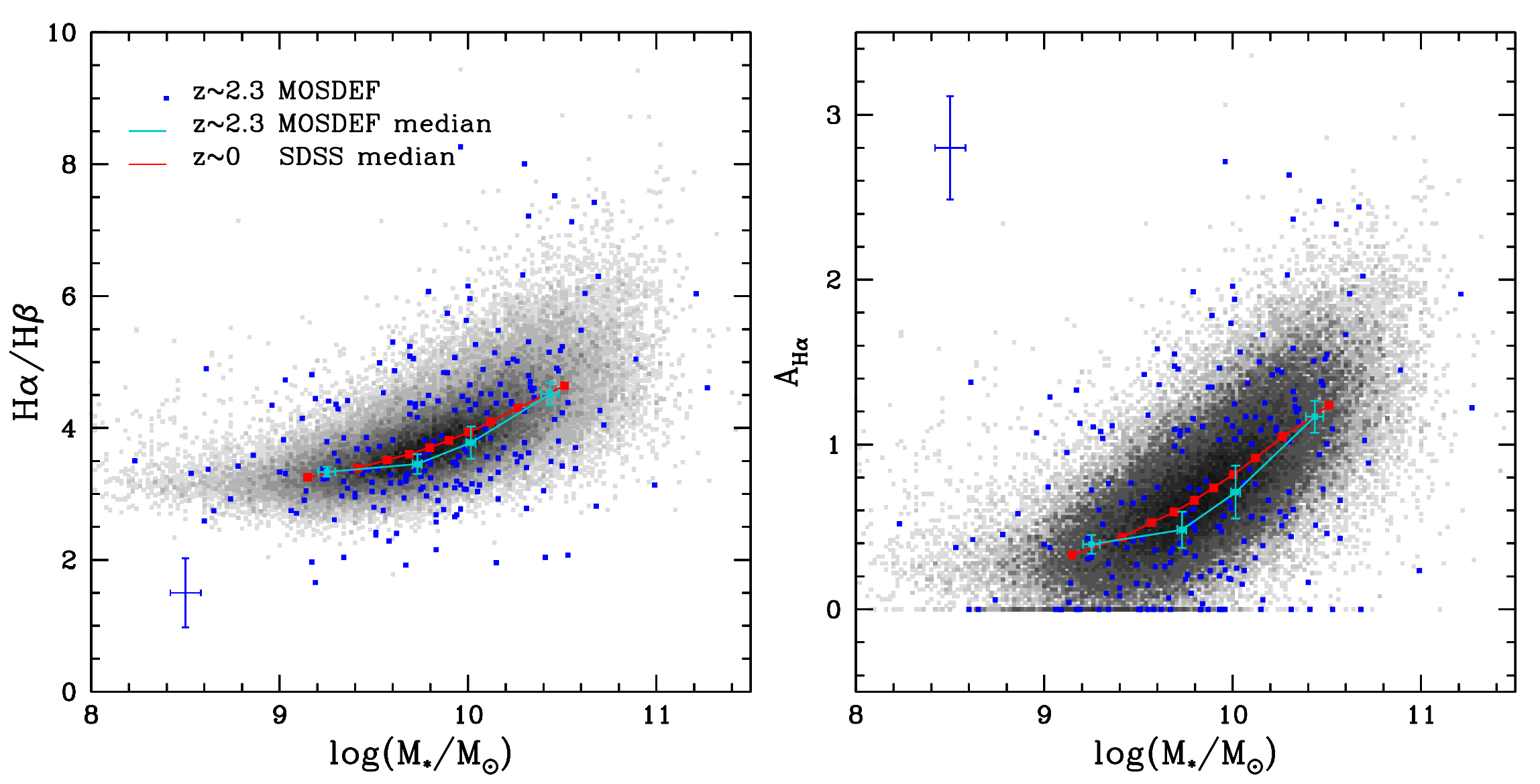}
\caption{Attenuation vs. $M_*$, based on the Balmer-line ratio, H$\alpha$/H$\beta$. In each panel,
$z\sim 2.3$ MOSDEF galaxies are indicated with blue points, and a median $z\sim 2.3$ error bar is shown
in the corner of the plot. The greyscale histogram corresponds to the distribution of local
SDSS galaxies. Running median H$\alpha$/H$\beta$ line ratios and the corresponding magnitude of
attenuation at the wavelength of H$\alpha$, $A_{{\rm H}\alpha}$,
are calculated in bins of stellar mass. The $z\sim 2.3$ running median is indicated in turquoise,
while that for SDSS is plotted in red. Median error bars are shown for the $z\sim 2.3$ sample,
while those for the SDSS sample are smaller than the symbols.
{\bf Left:} H$\alpha$/H$\beta$ ratio vs. $M_*$. {\bf Right:} $A_{{\rm H}\alpha}$ vs. $M_*$.
}
\label{fig:hahbpanels}
\end{figure*}

\subsection{SDSS Comparison Sample}
\label{sec:obs-sdss}
In order to perform an evolutionary comparison, we selected a sample of local galaxies from
the Sloan Digital Sky Survey (SDSS) Data release 7 \citep[DR7;][]{abazajian2009}.
Stellar masses and emission-line measurements corrected for stellar absorption were drawn from the MPA-JHU catalog of measurements for 
DR7\footnote{Available at http://www.mpa-garching.mpg.de/SDSS/DR7/}. 
In this catalog, SDSS stellar masses were estimated by fitting a grid of \citet{bruzual2003} models spanning
a wide range in star-formation histories to $u$, $g$, $r$, $i$, and $z$ emission-line-corrected photometry.
We restricted 
the SDSS sample to galaxies at $0.04 \leq z \leq 0.10$ to reduce aperture effects, and, following \citet{andrews2013}, required 5$\sigma$ detections
for [OII]$\lambda\lambda3726,3729$, H$\beta$, H$\alpha$, and [NII]$\lambda6584$, and a 3$\sigma$ detection for [OIII]$\lambda 5007$.
We also removed galaxies satisfying the optical emission-line AGN criterion of \citet{kauffmann2003}, yielding
a $z\sim 0$ comparison sample of 73,492 galaxies with $\log(M_*/M_{\odot})_{\rm med}=9.85$.
The SDSS sample is very well-matched to the $z\sim 2.3$ MOSDEF sample in terms of median stellar mass.
However, the median SFR for the SDSS sample is ${\rm SFR}_{\rm med} = 1.3 M_{\odot} \mbox{ yr}^{-1}$, i.e., a 
factor of $\sim 20$ lower, which reflects the evolution of the star-forming main sequence \citep[e.g.,][and references therein]{forsterschreiber2020}.
Metallicities for SDSS galaxies were estimated using the calibrations presented in Figure~3 of \citet{sanders2021}, in order
to provide the fairest comparison with respect to the $z\sim 2.3$ MOSDEF sample. 
Specifically, SDSS emission-line fluxes
were corrected for the contribution of diffuse ionized gas (DIG) following \citet{sanders2017}, and then the combination
of [OIII]$\lambda5007$/H$\beta$ and [OIII]$\lambda5007$/[OII]$\lambda\lambda3726,3729$ was fit simultaneously with the \citet{sanders2021} $z\sim 0$ DIG-corrected metallicity calibrations to yield $12+\log({\rm O/H})$.

In order to compare measures of rest-UV attenuation, we used the GALEX-SDSS-WISE Legacy catalog (GSWLC)
presented in \citet{salim2016}. As described in \citet{salim2016}, UV/optical galaxy SEDs including Galaxy Evolution Explorer (GALEX) 
and SDSS photometry were modeled
using the CIGALE code \citep{boquien2019}, including two-component exponentially-declining star-formation histories
generated using \citet{bruzual2003}, and a range of dust attenuation laws with varying UV slopes and UV-bump
strengths. Constraints on the preferred dust law were obtained by forcing agreement between SED-based and mid-IR
(from the Wide-field Infrared Survey Explorer; WISE) dust luminosities. One of the key outputs of the SED modeling using the GSWLC is the rest-UV attenuation, $A_{1600}$.
GALEX near-UV and far-UV coverage is available for 30,204 galaxies in our SDSS comparison sample. We have confirmed
that this GALEX subsample is representative of the larger SDSS-only sample, in terms of the relationship between Balmer line ratios and stellar mass (see Section~\ref{sec:results}).

\section{Results}
\label{sec:results}

The relationship between attenuation and stellar mass has been considered using
several different measures of the effects of dust. These include UV attenuation, $A_{1600}$ \citep{mclure2018},
rest-optical continuum and line attenuation, $A_V$ and $A_{{\rm H}\alpha}$ \citep{cullen2018,garn2010},
the ratio of FIR to UV luminosity, ``IRX" \citep{heinis2014,bourne2017}, the fraction of star formation
that is obscured, $f_{{\rm obscured}}$ \citep{whitaker2017}, and the Balmer decrement \citep{kashino2013,dominguez2013,price2014}. Here we analyze the attenuation of both
rest-optical lines and rest-UV continuum  as a function of stellar mass, and compare these relations
for star-forming galaxies at $z\sim 0$ and $z\sim 2.3$.

First, we consider attenuation estimated from the ratio of Balmer emission lines, H$\alpha$/H$\beta$.
Figure~\ref{fig:hahbpanels} (left) shows that there is a significant correlation between H$\alpha$/H$\beta$
and $M_*$, such that higher H$\alpha$/H$\beta$ ratios are associated with higher $M_*$. This correlation
applies to both the $z\sim 0$ SDSS and $z\sim 2.3$ MOSDEF samples. We also plot the running median 
of H$\alpha$/H$\beta$ in bins of $M_*$. The $z\sim 2.3$ running median relation between H$\alpha$/H$\beta$
and $M_*$ is entirely consistent with that of the $z\sim 0$ sample. The H$\alpha$/H$\beta$ ratio can be translated
into the nebular attenuation at the wavelength of H$\alpha$ using a specific dust extinction curve.
For star-forming galaxies in the local universe, the Milky Way curve of \citet{cardelli1989} is typically
used to interpret nebular reddening. Recently, \citet{reddy2020} demonstrated that the  \citet{cardelli1989}
curve is also appropriate for dust-correcting nebular emission lines in MOSDEF galaxies at $1.4 \leq z \leq 2.6$.
Using the \citet{cardelli1989} curve for both $z\sim 0$ and $z\sim 2.3$ samples, we cast the H$\alpha$/H$\beta$
vs. $M_*$ plot in terms of $A_{{\rm H}\alpha}$, the magnitude of nebular attenuation at the wavelength
of H$\alpha$ (Figure~\ref{fig:hahbpanels}, right). As in the case of H$\alpha$/H$\beta$, the running
medians of $A_{{\rm H}\alpha}$ in bins of $M_*$ are also consistent between $z\sim 0$ and $z\sim 2.3$.
We also note that the $z\sim 0$ relationship presented here between $A_{{\rm H}\alpha}$ and $M_*$ is entirely
consistent with that of \citet{garn2010}, when the same nebular attenuation law is assumed. Finally,
the stacked $z\sim 1.6$ $A_{{\rm H}\alpha}$ measurements of \citet{kashino2013}, in three bins of $M_*$, 
are consistent with both our $z\sim 2.3$ and the SDSS $z\sim 0$ median relations in Figure~\ref{fig:hahbpanels},
when the same stellar IMF and nebular attenuation law are assumed.

\begin{figure}[t!]
\centering
\includegraphics[width=0.95\linewidth]{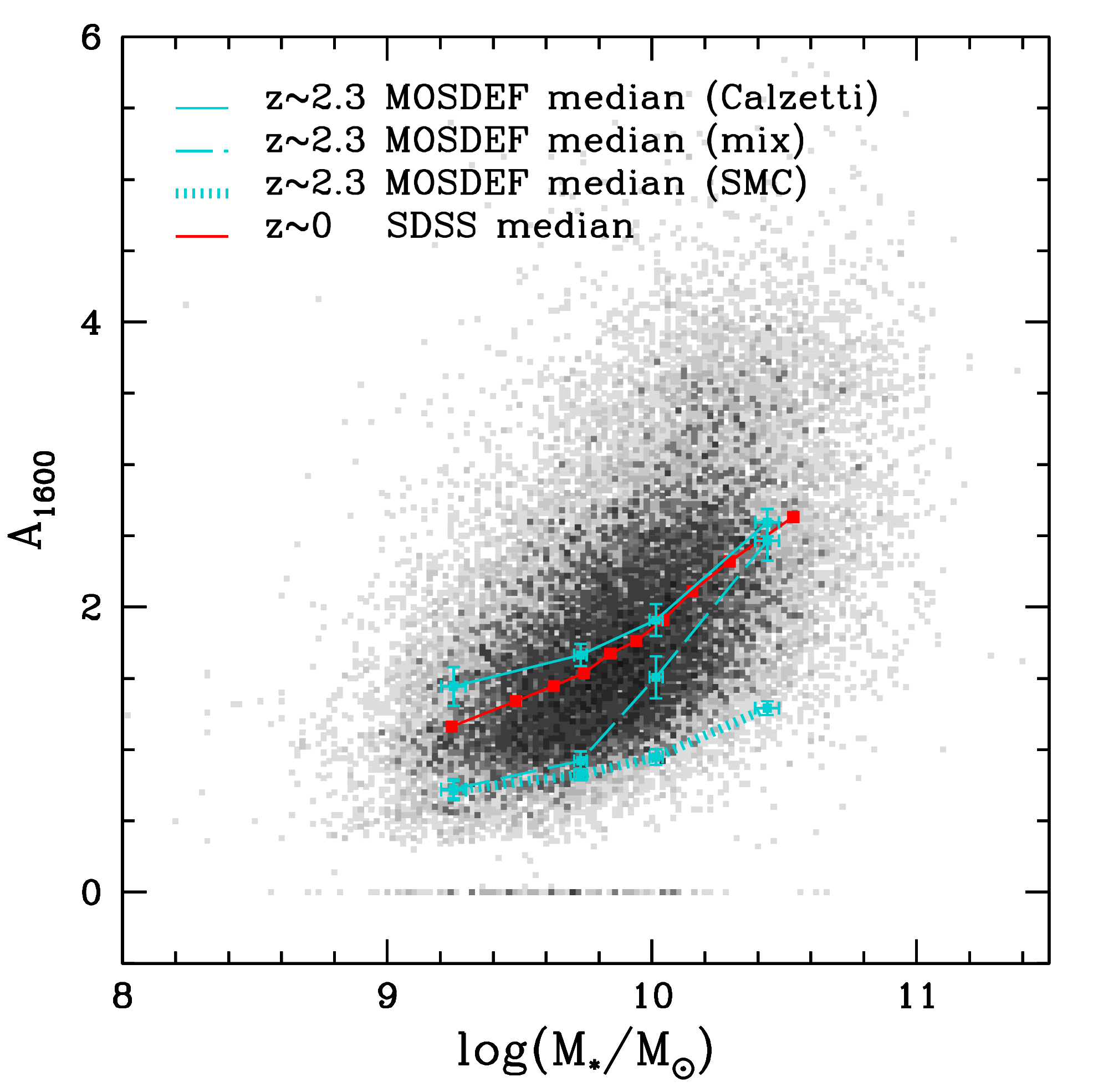}
\caption{Attenuation vs. $M_*$, based on the rest-UV continuum. Here attenuation is quantified as $A_{1600}$,
the magnitude of attenuation at a rest wavelength of $1600$\AA. As in Figure~\ref{fig:hahbpanels}, the local SDSS
sample is indicated as a grey histogram, and its running median $A_{1600}$ estimated in bins of $M_*$ is shown
with red, connected symbols. $A_{1600}$ is determined for SDSS galaxies as described in \citet{salim2016}.
The running median $A_{1600}$ for the $z\sim 2.3$ MOSDEF sample in bins of $M_*$ is shown in turquoise, using three different prescriptions
to translate the measured UV slope value, $\beta$, to $A_{1600}$. The solid curve shows $A_{1600}$ based on assuming
a \citet{calzetti2000} dust attenuation curve; the dashed curve shows $A_{1600}$ assuming a \citet{calzetti2000} curve at
$12+\log({\rm O/H})\geq 8.5$ and an SMC curve at $12+\log({\rm O/H})<8.5$; finally, the dotted curve shows $A_{1600}$
assuming an SMC curve for the entire MOSDEF $z\sim 2.3$ sample (a scenario not favored by the MOSDEF data, but included
for completeness). As in Figure~\ref{fig:hahbpanels}, median error bars are shown for the $z\sim 2.3$ sample,
while those for the SDSS sample are smaller than the symbols.
}
\label{fig:a1600lm}
\end{figure}

\begin{figure}
\centering
\includegraphics[width=0.95\linewidth]{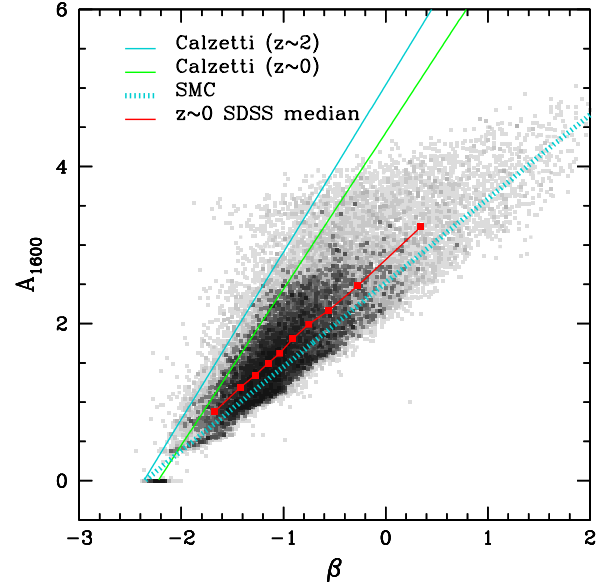}
\caption{UV Attenuation ($A_{1600}$) vs. UV slope $\beta$ for local SDSS galaxies. The SDSS sample
is indicated as a grey histogram, and its running median $A_{1600}$ estimated in bins of $\beta$ is shown
with red, connected symbols. Also shown as bracketing cases are the expressions relating $A_{1600}$ and $\beta$
from equations~(\ref{eq:a1600calz}) and (\ref{eq:a1600smc}). The solid,
turquoise line indicates the $A_{1600}$-$\beta$ relation assuming the \citet{calzetti2000} dust curve,
while the dotted turquoise line shows the $A_{1600}$-$\beta$ relation assuming the SMC dust curve.
In addition, we show with a green, solid line the $A_{1600}$-$\beta$ relation derived for $z\sim0$ starbursts by \citet{meurer1999}. This relation is based on a slightly redder intrinsic UV slope than the one we have assumed for $z\sim 2$ star-forming galaxies and is consistent
with the \citet{calzetti2000} curve for $z\sim 0$ starbursts.
The median relation between $A_{1600}$ and $\beta$ for SDSS galaxies, based on SED fitting and energy
balance, more closely resembles that for an SMC dust curve than either \citet{calzetti2000} relation, at least in the
UV regime.
}
\label{fig:a1600beta}
\end{figure}

Dust attenuation is also commonly quantified in terms of its effect on the rest-UV stellar continuum.
Such estimates of dust attenuation can be based on measurements of the UV slope, $\beta$, or else
from SED fitting over a wider wavelength range. As described in Section~\ref{sec:obs-mosdef},
for the $z\sim 2.3$ MOSDEF sample, we use measurements of $\beta$ and either the \citet{calzetti2000}
dust curve at all metallicities (and masses), or else a metallicity-dependent dust curve (\citealt{calzetti2000} at high metallicity
and SMC at low metallicity) to estimate $A_{1600}$. For the $z\sim 0$ SDSS sample, $A_{1600}$ is inferred
from UV/optical SED fitting and energy balance considerations. 

Figure~\ref{fig:a1600lm} shows
the relationship between $A_{1600}$  and $M_*$. For the sake of simplifying the plot while still
conveying key information, we do not display individual $z\sim 2.3$ datapoints under different assumptions
regarding the dust curve, but rather only the corresponding running medians. Over the full range in
stellar mass, the $z\sim 2.3$ and $z\sim 0$ medians in $A_{1600}$ are consistent within the errors
when the \citet{calzetti2000} curve is exclusively assumed (here $z\sim 2.3$ points are connected by a solid curve). 
In this case, the $z\sim 2.3$ median $A_{1600}$ value
fall slightly higher than their $z\sim 0$ counterparts at the same stellar mass. 
Our results are consistent with those of \citet{mclure2018}, equation (17), 
in the highest- and lowest-mass bins, though, on average those authors find slightly
higher $A_{1600}$ at fixed mass than we do here.
If we use a metallicity-dependent dust law to convert $\beta$ to $A_{1600}$ (here $z\sim 2.3$ points are connected by a dashed
curve), the $z\sim 2.3$ $A_{1600}$ median
values fall below the corresponding datapoints at $z\sim 0$ and fixed stellar mass. The $z\sim 2.3$
medians are consistent with those at  $z\sim 0$ down to a stellar mass of $\sim 10^{10} M_{\odot}$ (at the $\sim 0.4\sigma$
and $2.5\sigma$ level, respectively, for the highest- and second-highest $z\sim 2.3$ stellar mass bins).
In the two lower-mass bins, the $z\sim 2.3$ medians diverge downwards from the $z\sim 0$ median curve at
the $6.5-9.4\sigma$ level. For completeness, we also show the running median $A_{1600}$ vs. stellar mass
at $z\sim 2.3$ if the SMC dust curve is assumed for the entire sample \citep{reddy2018}, in which case the $z\sim 2.3$
sample running median $A_{1600}$ falls significantly below the $z\sim 0$ trend (here $z\sim 2.3$ points are connected by a dotted 
curve). However, we argue below
that this final set of assumptions is inconsistent with other previous MOSDEF results.
Based on the two more likely, bracketing,  prescriptions for converting $\beta$ to $A_{1600}$ at $z\sim 2.3$,
we also infer no significant evolution from $z\sim 0$ to $z\sim 2.3$
in the relationship between $A_{1600}$ and $M_*$.

It is worth emphasizing the distinct methodologies used for inferring $A_{1600}$ at 
$z\sim 0$ and $z\sim 2.3$, each of which based on the available information at that redshift.
In Figure~\ref{fig:a1600beta}, we show the relationship between $A_{1600}$ and $\beta$ for the $z\sim 0$ SDSS sample, where $\beta$
and $A_{1600}$ are inferred from SED fitting spanning from the UV to optical, along with longer-wavelength energy
balance considerations as described in \citet{salim2016}. Along with the running median  $A_{1600}$ for SDSS in
bins of $\beta$, we overplot the relations from equations~(\ref{eq:a1600calz}) and (\ref{eq:a1600smc})
as two bracketing cases, along with the $z\sim 0$ Calzetti relation from \citet{meurer1999}. The median SDSS relation falls in between those dictated by the \citet{calzetti2000}
and SMC curves, but significantly closer to the relationship traced by the SMC curve \citep[see also][]{salim2018}. 
Accordingly, while the SDSS sample is described by a wide range of dust attenuation curves \citep{salim2016}, 
on average these curves are steeper than the \citet{calzetti2000} curve, 
and similar in UV slope to the SMC curve \citep{salim2018}.

We do not have
direct $A_{1600}$ estimates for the majority of MOSDEF galaxies due to a lack of individual far-IR photometric
detections, and rely on $\beta$ measurements and
the {\it assumption} of different dust curves to infer $A_{1600}$. At least within the high-mass
($M_*\sim 10^{10}-10^{11}M_{\odot}$), high-SFR ($30-250 M_{\odot}\mbox{ yr}^{-1}$) 
range of the  $z\sim 2.3$ MOSDEF sample,
where {\it Spitzer}/MIPS, {\it Herschel}/PACS and SPIRE detections have been achieved,
SED fitting by \citet{shivaei2016} suggests that  the \citet{calzetti2000} (and not an SMC) law provides an accurate
description of the energy balance between rest-frame UV and far-IR. Similar dust-continuum measurements
are now required for lower-mass, lower-luminosity $z\sim 2.3$ MOSDEF galaxies, as in \citet{reddy2018}.

\section{Discussion}
\label{sec:discussion}
We find no significant evolution in the relationship between dust attenuation and stellar
mass between $z\sim 0$ and $z\sim 2.3$. Our results join a list of several that
arrive at similar conclusions, based on different proxies for dust attenuation 
($L_{IR}/L_{UV}$, or IRX; the fraction of obscured star formation, $f_{{\rm obscured}}$; 
$A_V$; and $A_{1600}$) and different multi-wavelength datasets \citep[e.g.,][]{heinis2014,pannella2015,bourne2017,whitaker2017,
mclure2018,cullen2018}. What is new here is the estimate of dust attenuation at $z\sim 2.3$ based
on a large set of individual Balmer decrement measurements, as well as a direct comparison of UV attenuation, $A_{1600}$,
at $z\sim 2.3$ and $z\sim 0$ based on GALEX data for the local SDSS comparison sample.
As we now discuss, the lack of strong evolution in the attenuation vs. stellar mass
relation has striking implications, based on what else is known about the evolution
of the ISM in star-forming galaxies between $z\sim 0$ and $z\sim 2.3$.

\subsection{The Connection between Attenuation and other ISM Properties}
\label{sec:disc-atten}

We start with the expression for dust attenuation at a given wavelength, $\lambda$.
For a given dust optical depth, $\tau_{\lambda}$, and attenuation $A_{\lambda}$,
with the latter in magnitudes, we find:

\begin{equation}
\exp(-\tau_{\lambda}) = 10^{-0.4 A_{\lambda}}
\label{eq:tauA}
\end{equation}

\noindent which corresponds to:

\begin{equation}
A_{\lambda}=\frac{\tau_{\lambda}}{0.4\ln(10)} = 1.086\times \tau_{\lambda}
\label{eq:Atau}
\end{equation}

We then recall the expression for $\tau_{\lambda}$, as a function
of the dust mass absorption coefficient, $\kappa_{\lambda}$, in units of m$^2$~kg$^{-1}$,
the dust density, $\rho_{{\rm dust}}$, in units of kg~m$^{-3}$, and the differential path length along the
line of sight, $ds$:

\begin{equation}
\tau_{\lambda}=\int \kappa_{\lambda} \times \rho_{{\rm dust}} ds
\label{eq:tauint}
\end{equation}

\noindent In a simplified model where $\kappa_{\lambda}$ is spatially independent and dust
is smoothly distributed throughout the galaxy disk, the integral can be re-expressed as the product of the dust mass absorption coefficient
and the dust mass surface density, $({M_{{\rm dust}}}/({\pi r_{{\rm dust}}^2}))$,
where $M_{{\rm dust}}$ is the dust mass and $r_{{\rm dust}}$ is the scale of the disk
over which dust is distributed:

\begin{equation}
\tau_{\lambda}\simeq\kappa_{\lambda}\times (\frac{M_{{\rm dust}}}{\pi r_{{\rm dust}}^2})
\label{eq:sigmadust}
\end{equation}

\noindent Folding in equation~(\ref{eq:Atau}), we can express $A_{\lambda}$ as a function
of dust properties:

\begin{equation}
A_{\lambda}\simeq 1.086\times \kappa_{\lambda}\times(\frac{M_{{\rm dust}}}{\pi r_{{\rm dust}}^2})
\label{eq:Asigmadust}
\end{equation}

\noindent Within the context of the simplified model presented here, the lack of significant evolution in $A_{\lambda}$ at fixed stellar mass from $z\sim 0$ to $z\sim 2.3$
implies that the product $\kappa_{\lambda}\times ({M_{{\rm dust}}}/({\pi r_{{\rm dust}}^2}))$
remains constant at fixed stellar mass. 

\subsubsection{Direct $M_{\rm dust}$ Measurements}
\label{sec:disc-atten-dust}
First, we consider what is known from direct measurements of dust masses at high redshift.
The Atacama Large Millimeter/submillimeter Array (ALMA) is now beginning to enable $M_{{\rm dust}}$
estimates for galaxies in the luminous end of the Luminous Infrared Galaxy (LIRG) 
regime (i.e., $10^{11} L_{\odot} \leq L_{\rm IR} \leq 10^{12} L_{\odot}$; \citealt{aravena2020}; Shivaei et al. 2021,
in prep), but does not yet cover the full range of stellar masses and SFRs in our sample.
However, initial ALMA results from \citet{magnelli2020} and \citet{donevski2020} 
indicate signficant evolution in
the $({M_{{\rm dust}}}/{M_*})$ ratio at fixed $M_*$ from the local universe out to $z\sim 2$. 
Based on stacked measurements of galaxies
covered by the ALMA Spectroscopic
Survey (ASPECS) large program,  \citet{magnelli2020} find a best-fit factor of 10 evolution in $({M_{{\rm dust}}}/{M_*})$
for star-forming galaxies at a fiducial stellar mass of $M_*=10^{10.7} M_{\odot}$ between $z=0.45$ and $z=2.0$.
Likewise, for star-forming galaxies individually detected by ALMA, at a median redshift of $z_{\rm med}=2.39$ and median
stellar mass of $M_*=10^{11}$, \citet{donevski2020} find an order of magnitude increase in $({M_{{\rm dust}}}/{M_*})$
relative to that observed in the local galaxy sample of \citet{andreani2018}.

This observed evolution in $({M_{{\rm dust}}}/{M_*})$
exists in tension with model predictions from, e.g., \citet{popping2017},
which include a roughly constant relationship between $M_{{\rm dust}}$ and $M_*$ over
the redshift range $z\sim 0$ to $z\sim 2$. As highlighted by \citet{popping2019}, such models also significantly (by a factor of $2-3$) underpredict the molecular gas content of $z\sim 2$ galaxies, which likely contributes to the discrepancy between their predicted and observed dust masses. The tension between the predicted and observed evolution of ISM gas and dust masses needs to be addressed.

\subsubsection{The $({M_{{\rm dust}}}/{M_{{\rm gas}}})$ Ratio and Gas Surface Density}
\label{sec:disc-atten-gas}
We can also consider the question of evolution in $A_{\lambda}$ by re-expressing the dust mass surface density,
$ ({M_{{\rm dust}}}/({\pi r_{{\rm dust}}^2}))$, i.e., $\Sigma_{\rm dust}$,
as the product $({M_{{\rm dust}}}/{M_{{\rm gas}}}) \times \Sigma_{\rm gas}$. Here, $\Sigma_{\rm gas}$
is the gas surface density. Accordingly, we can rewrite equation~(\ref{eq:Asigmadust}) as:

\begin{equation}
A_{\lambda}\simeq 1.086\times \kappa_{\lambda}\times (\frac{M_{\rm dust}}{M_{\rm gas}})\times \Sigma_{\rm gas}
\label{eq:Asigmagas}
\end{equation}

\noindent With this equation, we can gain indirect constraints on the evolution of $\Sigma_{\rm dust}$ from 
estimates of the evolution of galaxy metallicity and gas surface density.  
For this analysis, we consider the evolution in galaxy properties
at a fiducial stellar mass of $10^{10} M_{\odot}$, close to the median stellar
mass of both the $z\sim 2.3$ MOSDEF and $z\sim 0$ SDSS samples.

\begin{figure}[t!]
\centering
\includegraphics[width=0.95\linewidth]{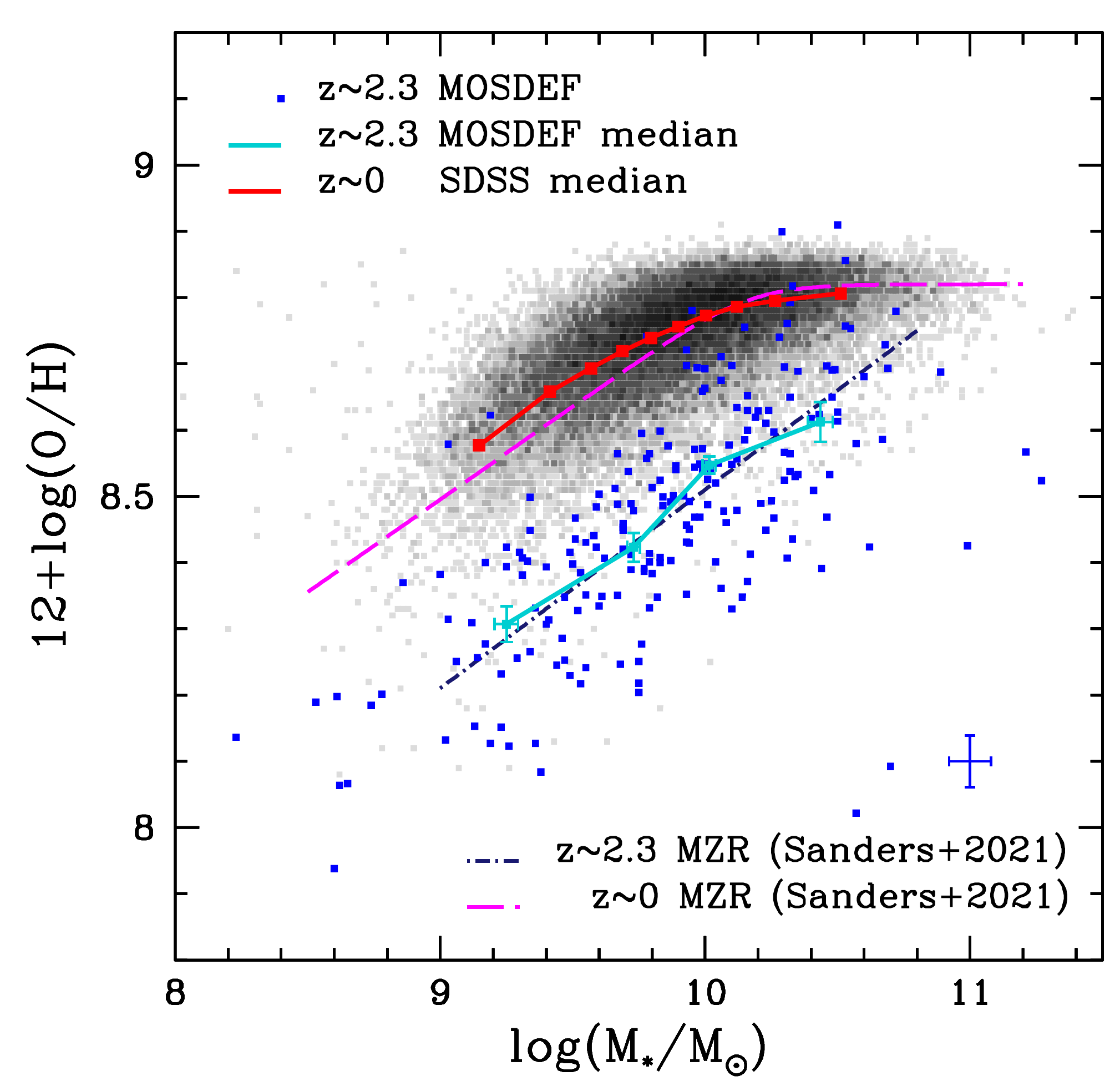}
\caption{$12+\log({\rm O/H})$ vs. $M_*$, i.e., the mass-metallicity relation (MZR).
Symbols and running medians for $z\sim 2.3$ MOSDEF galaxies and local SDSS galaxies are as in Figure~\ref{fig:hahbpanels}.
$12+\log({\rm O/H})$ is estimated for $z\sim 0$ and $z\sim 2.3$ galaxies following the prescriptions in \citet{sanders2021}.
Also plotted are the best-fit $z\sim 0$ and $z\sim 2.3$ MZR relations from Table~3 of \citet{sanders2021}. The $z\sim 2.3$
best-fit relation exactly follows the running median $12+\log({\rm O/H})$ in bins of $M_*$, while the $z\sim 0$
best-fit relation falls slightly below our running median at $\log(M_*/M_{\odot})<10$, due to small differences
in SDSS comparison sample selection between the two works. At $M_*\sim 10^{10} M_{\odot}$, however, there is
agreement between the best-fit SDSS MZR and our corresponding running median value. There is clear
evolution towards lower $12+\log({\rm O/H})$ at fixed $M_*$, which should be accompanied by a lower
$({M_{{\rm dust}}}/{M_{{\rm gas}}})$ according to the $({M_{{\rm dust}}}/{M_{{\rm gas}}})$ vs. $12+\log({\rm O/H})$
relation in \citet{devis2019}.
}
\label{fig:ohlm}
\end{figure}

In the first step, we can use the evolution in metallicity from 
$z\sim 0$ to $z\sim 2.3$ to infer the evolution in $({M_{{\rm dust}}}/{M_{{\rm gas}}})$.
As shown in Figure~\ref{fig:ohlm} and previous works \citep[e.g.,][]{steidel2014,
sanders2015,sanders2018,sanders2021}, $z\sim 2.3$ galaxies have lower metallicity
at fixed stellar mass relative to galaxies at $z\sim 0$. In addition to the local
and $z\sim 2.3$ samples analyzed here, we plot the best-fit mass-metallicity relations
from \citet{sanders2021}, which are consistent with our measurements and show
that our dust attenuation sample is representative of MOSDEF star-forming galaxies.
The offset in metallicity is $\Delta 12+\log({\rm O/H})=0.26\pm0.02$~dex towards lower
metallicity at $z\sim 2.3$. This difference in metallicity can be translated into
a reduction in $({M_{{\rm dust}}}/{M_{{\rm gas}}})$ from $z\sim 0$ to $z\sim 2.3$
via the relationship between $({M_{{\rm dust}}}/{M_{{\rm gas}}})$ and $12+\log({\rm O/H})$.
Most recently, \citet{devis2019} constructed this relationship for a large sample of local
galaxies with $M_{\rm dust}$, $M_{\rm gas}$, and $12+\log({\rm O/H})$ measurements.
While the slope, $a$, of the relationship 

\begin{equation}
\log (\frac{M_{{\rm dust}}}{M_{{\rm gas}}}) = a \times [12+\log({\rm O/H})] + b
\label{eq:devis2019}
\end{equation}

\noindent varies in detail depending on which empirical strong-line
metallicity indictator is adopted, Table~4 of \citet{devis2019} shows that the
values for $a$ are distributed around $2$. \citet{shapley2020} reported
evidence for a lack of evolution in the $\log ({M_{{\rm dust}}}/{M_{{\rm gas}}})$ 
vs.  $12+\log({\rm O/H})$ relationship, at least at the high-mass
end ($10^{10.5} M_{\odot} \lesssim M_* \lesssim 10^{11} M_{\odot}$), and, therefore, we assume that the local relation
can be applied to the $z\sim 2.3$ MOSDEF sample. Accordingly, we adopt a 
slope of $a=2.0\pm0.2$ for the $({M_{{\rm dust}}}/{M_{{\rm gas}}})$
vs. $12+\log({\rm O/H})$  relationship, and find that the decrease in 
$12+\log({\rm O/H})$ implies a decrease of $0.52\pm0.05$~dex in $({M_{{\rm dust}}}/{M_{{\rm gas}}})$
at fixed mass (i.e., a linear factor of $3.3\pm0.4$).

Next, we consider evolution in the gas surface density of star-forming galaxies out
to $z\sim 2.3$. The molecular gas surface density evolves dramatically at fixed stellar mass
over this redshift range. This evolution can be traced either by direct CO measurements \citep{tacconi2013}, or else measurements of the change
in SFR surface density, $\Sigma_{\rm SFR}$, coupled with an inversion of the Kennicutt-Schmidt (K-S) Law. For example,
based on the dust-corrected H$\alpha$ SFRs and rest-optical half-light radii for both our $z\sim 2.3$ MOSDEF and $z\sim 0$
SDSS comparison samples, we find an increase of a factor of $\sim 65$ in the median $\Sigma_{\rm SFR}$
\citep[see also][]{shapley2019}.  Assuming a linear power-law slope for the K-S law \citep{tacconi2013}, we infer an increase
in molecular $\Sigma_{\rm gas}$ by the same factor of 65 from $z\sim 0$ to $z\sim 2.3$. While the gas content of
$z\sim2$ star-forming galaxies is well-approximated by the molecular component \citep{tacconi2018},
in local galaxies at the median stellar mass of our SDSS sample the molecular component only comprises $\sim 20$\%
of the total molecular plus atomic gas mass \citep[][Table 3]{catinella2018}. The evolution in total
$\Sigma_{\rm gas}$ from $z\sim 0$ to $z\sim 2.3$ is therefore reduced by a factor of $\sim 5$ relative to
the inferred evolution in molecular $\Sigma_{\rm gas}$, since the total
$\Sigma_{\rm gas}$ for $z\sim 0$ galaxies is a factor of $\sim 5$ higher than the molecular component alone.
Total $\Sigma_{\rm gas}$ is thus inferred to increase by a factor of $\sim 13$.
The net effect of the inferred evolution in $({M_{{\rm dust}}}/{M_{{\rm gas}}})$
and $\Sigma_{\rm gas}$ is a factor of $({M_{{\rm dust}}}/{M_{{\rm gas}}})\times \Sigma_{\rm gas} \sim (1/3.3)\times 13$,
i.e., a factor of $\sim 4$. This factor corresponds to the increase in $\Sigma_{\rm dust}$, assuming that the spatial
extent of dust and molecular gas is the same. ALMA has been used to obtain spatially-resolved dust-continuum and
CO maps for small samples of massive ($M_*>10^{11} M_{\odot}$) and
luminous ($L>10^{12} L_{\odot}$) galaxies at $z\sim 2$ \citep{tadaki2017,calistrorivera2018,kaasinen2020}, but a clear picture has
yet to emerge from these measurements regarding the relative extents of dust and molecular gas emission. A larger 
sample of spatially resolved measurements
of both dust continuum and CO emission is required for less luminous $z\sim 2.3$ main sequence galaxies in the LIRG regime in order to 
understand if the extent of dust continuum emission evolves in the same manner as that of the molecular gas.

\subsection{Explaining the Lack of Evolution}
\label{sec:disc-explanation}
In the extremely simplified picture presented in Section~\ref{sec:disc-atten-dust}, in 
order to maintain a fixed attenuation $A_{\lambda}$ at fixed
stellar mass there must be evolution in either $\kappa_{\lambda}$, $r_{{\rm dust}}$, or both,
in the sense that $\kappa_{\lambda}$ is {\it smaller} and $r_{{\rm dust}}$ is effectively
{\it larger} at $z\sim 2.3$ than at $z\sim 0$. Alternatively (or in addition), following the discussion
in Section~\ref{sec:disc-atten-gas}, if there is a stronger dependence
of $({M_{{\rm dust}}}/{M_{{\rm gas}}})$ on metallicity at $z\sim 2.3$ than
at $z\sim 0$ \citep{devis2019}, such that a reduction of $0.26$~dex in metallicity corresponds
to a more extreme decrease in $({M_{{\rm dust}}}/{M_{{\rm gas}}})$, this effect
would also help to explain the lack of evolution in $A_{\lambda}$ at fixed stellar
mass. While additional data at lower
metallicities is required to show the actual form of the relation at $z\sim 2.3$,
initial results from \citet{shapley2020} suggest that the normalization
in the $({M_{{\rm dust}}}/{M_{{\rm gas}}})$ vs. $12+\log({\rm O/H})$
relation does not evolve at solar metallicity.

We now consider one of the first two factors: the spatial extent of dust, $r_{\rm dust}$.
Evolution in $r_{\rm dust}$ comprises another possibility for explaining the constant attenuation vs. stellar mass
relation in the face of significant $({M_{{\rm dust}}}/{M_*})$ evolution.
If ${\pi r_{{\rm dust}}^2}$ is a factor of $\sim 10$ larger at $z\sim 2.3$ at fixed $M_*$, corresponding
to an increase of a factor of $\sim 3$ in $r_{\rm dust}$, then $({M_{{\rm dust}}}/({\pi r_{{\rm dust}}^2}))$
would remain constant. However, an evolution towards larger $r_{\rm dust}$ at higher redshift and fixed stellar
mass is in conflict with both recent numerical simulations of galaxy formation including dust
radiative transfer \citep{popping2021}, as well as preliminary resolved ALMA measurements
of the evolution of dust sizes. Specifically, \citet{fujimoto2017} show for a sample of luminous ($L_{\rm FIR}\geq 10^{12}$)
and massive ($\log(M_*/M_{\odot})_{\rm med}\sim 11$) galaxies drawn from the DANCING-ALMA survey,
that rest-frame far-IR sizes measured with ALMA (tracing dust-continuum)
decrease over the range $1.5 \leq z \leq 4$. \citet{gomezguijarro2021}
find a similar evolution towards smaller rest-frame far-IR continuum sizes as redshift increases over $z\sim 2-4$,
based on GOODS-ALMA-2.0, a blind survey conducted at 1.1~mm covering a similar luminosity range. For a sample
of four $z\sim 1.5-2.0$ galaxies with lower
far-IR luminosities, in the LIRG range, \citet{cheng2020} finds comparable $r_{\rm dust}$  values to those of
local LIRGS from the KINGFISH \citep{kennicutt2011} and GOALS \citep{armus2009} surveys. However, there is no evidence 
to date for {\it larger} $r_{\rm dust}$ at higher redshift. 

It may be that simply using an effective size, ``$r_{\rm dust}$,"
is insufficient to capture differences in the typical spatial distributions of 
dust at $z\sim 0$ and $z\sim 2.3$. For example, if dust distributions are 
patchier and clumpier at $z\sim 2$ than locally, the observed dust attenuation for a 
given $M_{\rm dust}$ will be lower \citep{witt2000,seon2016}.
Spatially-resolved maps of dust-continuum emission extending up to $z\sim 2$ will
be crucial for addressing this question. In addition, a detailed analysis of the shapes
of dust attenuation curves for galaxies of similar mass at low and high redshift can be used
to determine the dust geometry indirectly in such systems \citep[e.g.,][]{chevallard2013}.

One final possibility for explaining the constant attenuation vs. stellar mass relation
consists of evolution in $\kappa_{\lambda}$, the wavelength-dependent dust mass absorption coefficient.
This dust cross-section per unit dust mass enscapsulates many different dust properties, including
dust-grain size distribution, grain morphology, density, and chemical composition. Recently,
\citet{clark2019} empirically determined maps of $\kappa_{\lambda}$ in two nearby face-on spiral
galaxies. \citet{clark2019} not only found significant variation of $\kappa_{\lambda}$ within the individual galaxies
targeted \citep[see also][]{bianchi2019}, but also that $\kappa_{\lambda}$ is {\it inversely} correlated with gas surface
density. Such an anti-correlation is not predicted by standard dust models, in which
denser ISM regions are conducive to the growth of larger grains, which have higher emissivity
(i.e., $\kappa_{\lambda}$) per unit mass. However, if lower $\kappa_{\lambda}$ is generally associated
with higher $\Sigma_{\rm gas}$, the significantly higher $\Sigma_{\rm gas}$ values at $z\sim 2.3$
described above may result in a lower $\kappa_{\lambda}$ for these high-redshift galaxies than their
low-redshift counterparts.

In summary, the roughly constant relationship between attenuation and stellar
mass from $z\sim 0$ to $z\sim 2.3$ poses an important puzzle, given the significant
evolution in the gas and dust content of the ISM at fixed stellar mass over
the same redshift range. We have highlighted multiple possiblities for explaining
the lack of evolution in attenuation vs. $M_*$. In particular, these include a steeper relationship
between $({M_{{\rm dust}}}/{M_{{\rm gas}}})$ and $12+\log({\rm O/H})$ at $z\sim 2.3$, such that
the decrease in $12+\log({\rm O/H})$ translates into lower $({M_{{\rm dust}}}/{M_{{\rm gas}}})$ at $z\sim 2.3$
than in the local universe; more extended dust distributions at $z\sim 2.3$ for galaxies at fixed stellar mass
(though such a possibility seems inconsisent with both theory and preliminary observations), or, on the other hand, {\it clumpier} dust distributions at $z\sim 2.3$; and 
a lower dust mass absorption coefficient $\kappa_{\lambda}$. Directly measuring $\kappa_{\lambda}$
at $z\sim 2.3$ seems beyond the reach of current facilities. However, determining 
$({M_{{\rm dust}}}/{M_{{\rm gas}}})$ vs. $12+\log({\rm O/H})$ at subsolar metallicities,
and obtaining spatially-resolved maps of the dust continuum emission for such galaxies
is well within the scope of ALMA. Such observations should be highly prioritized in order to
solve the puzzle of the non-evolving attenution vs. stellar mass relation.

\section*{Acknowledgements}
We acknowledge support from NSF AAG grants AST-1312780, 1312547, 1312764, 1313171, 2009313, and 2009085, grant AR-13907
from the Space Telescope Science Institute, grant NNX16AF54G from the NASA ADAP program,
and the support of the UK Science and Technologies Facilities Council.
Support for this work was also provided through the NASA Hubble Fellowship
grant \#HST-HF2-51469.001-A awarded by the Space Telescope
Science Institute, which is operated by the Association of Universities
for Research in Astronomy, Incorporated, under NASA contract NAS5-26555.
We acknowledge helpful conversations with Ian Smail, Natascha F\"orster
Schreiber, Tim Heckman, and John Peacock.  We finally wish to extend special thanks to those of Hawaiian ancestry on
whose sacred mountain we are privileged to be guests. Without their generous
hospitality, the work presented herein would not have been possible.


\end{document}